# AI coach for badminton

Dhruv Toshniwal
*School of Computer Engineering and Technology*
*Dr. Vishwanath Karad MIT World Peace University*
Pune, India
dhruvajaytoshniwal@gmail.com

Arpit Patil
*School of Computer Engineering and Technology*
*Dr. Vishwanath Karad MIT World Peace University*
Pune, India
arpitpatil26@gmail.com

Nancy Vachhani
*School of Computer Engineering and Technology*
*Dr. Vishwanath Karad MIT World Peace University*
Pune, India
nancyvachhani0@gmail.com

*Abstract*— In this competitive world of sports, to become an ideal player, one must maintain his nutrition and physique. Every motion involved in the sport must be efficiently carried out to employ the active muscles and eventually sustain energy depletion levels. Badminton is one sport where player movements are relatively easier to track using video analytics. The stroke movements captured through the camera can be used to examine the hand and hip coordination, placement of the leg and the varied upper bound and lower bound angles of the strokes. The prediction of a much better stance and orientation of the muscles, erroneous playing techniques, joint fatigue and several such aspects can be forecasted through an intelligent system as a recommendation. This research focuses on various techniques of neural networks to analyse the image retrieved from a badminton game. Ideal and efficient data is vastly available on the World Wide Web, and a series of data is then be taken down as per the requirements of the subjects, i.e., taking into consideration, player's body type and consecutively keeping the subjects diet and physique under observation while noting down the muscles which get activated during the execution of the said stroke.

## I. Introduction

### I.I. What is sports analysis?

'What' is the necessity for Sports Analysis?' is the answer to the explanation of Sports Analysis. Every professional athlete needs a coach who keeps track of their fundamental lifestyle. A coach can help an athlete minimize the chances of defects they make to improve their performance. When a coach examines an athlete during a competition, this is referred to as sports analysis. The athlete's efficiency, endurance, and performance should be prioritised. The coaches' goal is to ensure that the athlete is focused, competent, and sound. Unfortunately, there will always be room for error. In this case, we consider the range for error will be due to the coach being human. Thus there will always be a spectrum of faults that we cannot transcend, i.e., the potential of an error in the human misunderstanding of perfection or any human error for that matter. This inefficiency can be reduced to zero when an Artificial Intelligence (AI) model is used instead of a human coach. The AI model will always be detailed, precise, and to the point, considering all of the limits necessary for the athlete's overall development. This is regarded as Sports Analysis.

The purpose of this study is to discuss several Neural Network algorithms for processing images extracted from real-time badminton games. Section II summarises the related work done in this area. Section III describes the numerous components of Sports Analysis, how they operate, and the design and flow of the suggested model. Section IV contains the conclusions of the article.

### I.II. Application in Badminton

This section addresses the 'how' of Sports Analysis. Neural network architectures might be used to provide sports analysis. There is an infinite amount of data, training data set, and testing dataset on the World Wide Web. The Olympics dataset is used as the training data set, and it is the cream of the crop since athletes represent their countries at the international level. Now, for the testing data set, the athlete who wants to enhance their capabilities records the game they play while simultaneously working on the flaws identified by the algorithm after the initial evaluation. The player must improve their nutrition, training, and sleeping pattern for maximum efficiency. The athlete should be confident and focused during the software's successive assessments.

When the algorithm produces a list of mistakes after the initial evaluation, the athlete should try to minimise these errors. The evaluation may include but is not limited to inappropriate posture, incorrect racket handling, incorrect range of lower and upper boundaries of the angle of the stroke, poor hip movement, leg movement, and more. After working on them and completing the second assessment, it should be highlighted that there should be a significant difference between the first and second testing results. Between the two evaluations, there must be a significant difference in the athletes' total endurance, efficiency, performance, and strength. This may also be used to forecast injuries and identify talent.

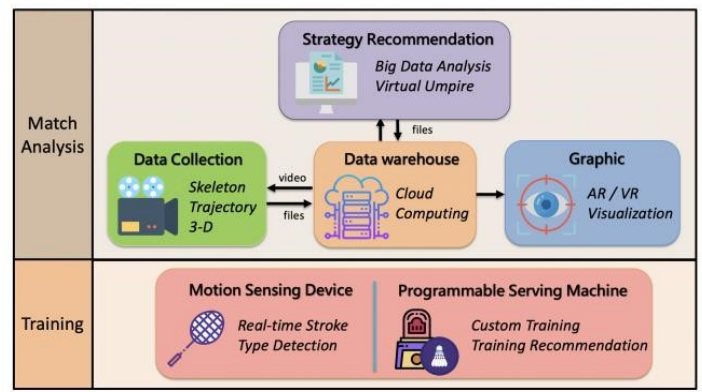

Figure 1. Scope of sports analysis (Badminton)

## II. LITERATURE REVIEW

As stated previously, numerous relevant studies have been done in this field, but not nearly enough. There is no actual, practical application for developing an interface for usage in the realm of badminton. The popularity of a field typically determines financing and the need for growth. Team sports such as cricket, badminton, ice hockey already have many applications in development, and many are in use. The works done by other researchers have established their credibility through a practical approach and also provided a gateway of the theoretical avenue for the model to be developed to any eager minds working in this particular field.

To comprehend sports analysis for badminton, one must first grasp the mechanics of the game and the necessity for an indoor court to make it an enclosed recreation. Why and how humidity, airflow, temperature, and wind direction may impact the game. Everything required has been incorporated in the works and is discussed further below.

### II.I TACTICAL ANALYSIS

Object tracking based on computer vision has been used extensively to supplement and augment sports footage [1]. There is no need to manually collect data because it will be retrieved systematically and automatically, and strategically examined. This comprises data visualisation, auxiliary device training, and data warehousing. As previously explained, several deep learning approaches can be used to construct video-based real-time microscopic competition information gathering based on broadcast tournament recordings. Tactical analysis has been done using machine learning techniques. Intelligent badminton rackets and wirelessly linked serving machines will be created as training auxiliary devices employing IoT technology to increase tactical data analysis and training efficiency.

Algorithms used include: OpenPose [6], TrackNet, tactical analysis, deep learning, machine learning, YOLOv3 [11], visualisation, cloud service, smart racket, programmable serving machine.

Also, a project for badminton competition data collection and tactical analysis has been introduced [1]. Deep learning techniques are adopted to develop video-based competition data collection, including shuttlecock trajectory and players' positions and skeletons. Machine learning techniques are used to develop tactical analysis based on microscopic data. Visualisation techniques are used to meaningfully and adequately represent and display microscopic and macroscopic competition data for easy understanding. Besides, training auxiliary devices, including intelligent badminton rackets and connected serving machines, will be developed to improve training efficiency. In the future, besides developing the techniques depicted in the roadmap, there is scope for developing the capability of 3D data collection and analysis and extending the research on singles matches to doubles matches.

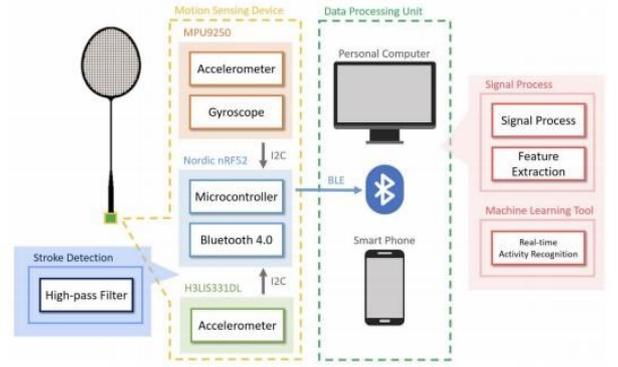

Figure 2. Architecture of a smart racket

### II.II. PHYSICS BEHIND BADMINTON

The aerodynamics of a shuttlecock's trajectory are detailed [2]. It flips like a light and elongated particle on a pure drag trajectory. Following a study of the flip phenomena and flight dynamics and a discussion of the game's ramifications, a possible categorization of distinct shots is presented. Because of changes in humidity, air pressure, temperature, and other factors, badminton is played indoors.

The most anticipated conclusion is identifying the benchmark value of critical factors that influenced the player's skill performance; object assessment is also projected in this project. Future studies should result in a complete training model for improving player skills. Other tests focusing on Badminton stroke-like smash, clear, lift, and service will be undertaken across a broader population of sample participants ranging from casual to professional level badminton players.

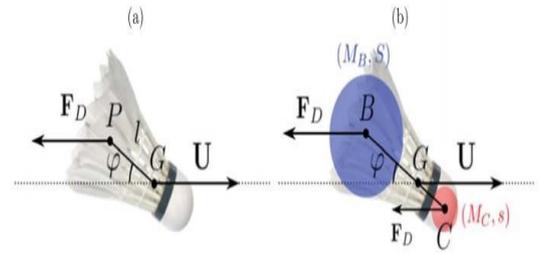

Figure 3. Drag force FD applied on a shuttlecock whose direction forms an angle φ with the velocity U. (b) Model system composed of a sphere of large section S and mass MB, which stands for the skirt, and a sphere of small section s and large mass MC, which represents the cork.

### II.III. SENSOR-BASED MODEL

Badminton is one of Malaysia's most popular sports, and the key objective of this research article is to examine the sets of different motions in badminton training using sensors to identify the ideal movement that optimizes badminton performance. Determine the measurable parameters that will aid in quantifying badminton skill levels [3]. The performance of elite players is examined to establish the benchmark values for these quantitative criteria. The factors are then used to develop a quantitative model that may be used to estimate performance levels. It will aid badminton players in improving their techniques and would give an objective assessment for grading badminton talents.

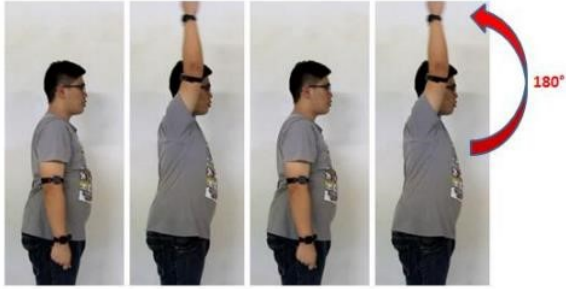

Figure 4. Flow of swing hand (180 degrees)

The most likely conclusion of this research stage is to establish the benchmark value of crucial variables that affect the player's skill performance; an objective assessment is required in this methodology. Further in-depth research and investigation of a sensor-based quantitative model for badminton skill analysis and assessment are necessary. A complete training model to develop player skills is expected in a future study. Other trials focusing on badminton strokes like smash, clear, lift, and service will be undertaken across a broader group of sample subjects, from casual to professional level Badminton players. Future studies should detect a player's position on the game court to develop a more comprehensive automated Badminton skill evaluation system.

## II.IV. MACHINE LEARNING ASSESSMENT

The Badminton assessment process is a method that evaluates players' performance and identifies their strengths and flaws in order to increase training effectiveness [4]. Several traditional evaluation approaches, which, as previously indicated, suffer from a lack of personnel, experience, and the fundamental objective model. The principal objective is to design and create an innovative and effective badminton assessment system. This approach employs three evaluation modules (Module 1: Badminton Serving Accuracy, Module 2: Badminton Shots Quality, and Module 3: Player's Agility) to extract the necessary quantitative metrics from players' serves, hits, and agility. In this work, kinematic characteristics such as acceleration, power, and rotational speed are collected using a 9-degree freedom wireless sensor, an APDM Opal sensor, and a badminton feedback sensor, XiaoYu 2.0. All three modules were verified with three strong and six average players, and 46 features were gathered. Using the t-test approach, 39 out of 46 characteristics were substantially different. Relief, Principal Component Analysis, and Correlation Feature Selection are the three feature extraction methodologies.

The obtained datasets are examined using seven machine learning models: Random Tree (RT), Random Forest, Artificial Neural Network, K Star, Multiple Linear Regression, Gaussian Process, and Support Vector Machine. There were a total of 21 rating models built. According to the data, the RT model has a prediction accuracy of 90.84% and a correlation value of r=0.86.

## II.V. VIDEO ANALYSIS AND BREAKDOWN OF BROADCAST BADMINTON VIDEOS

For a long time, sports analytics has piqued the curiosity of the computer vision community. Video summary, highlight generating, coaching help, player fitness, flaws and strengths evaluation, and so on are all applications of sport analytic systems. Sports videos intended for live watching are usually accessible in broadcast videos for consumers. Several thousand hours of broadcast video are already available on the internet. Sports broadcast footage are frequently lengthy and shot in a natural atmosphere from diverse angles. These clips are frequently edited, and extra theatricals, such as animations or graphics, are included.

A limited technique of automated annotation and informative analytics of sports broadcast films, specifically badminton games, exists [5]. Like many other sports, badminton has a distinct game grammar; a well-separated playing court is structured as a succession of events (points, rallies, and winning points) and lends itself well to large-scale analytics. The detection of players, points, and strokes for each match frame allows for quick indexing and retrieval. Furthermore, these precise annotations generate clear metrics for higher-level analytics.

Such systems have several advantages. While quantitative scores highlight players' performance, qualitative game analysis improves the viewing experience. The player's tactics, skills, and weaknesses might be mined and easily highlighted for training. It automates several aspects of analysis traditionally carried out manually by specialists and coaches.

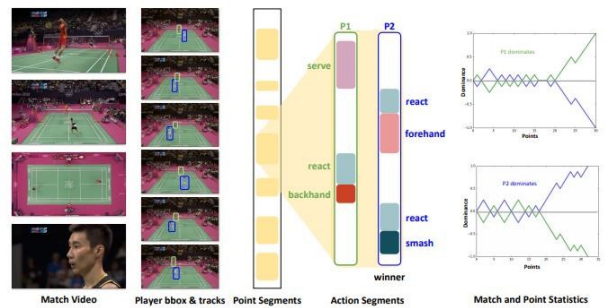

Figure 5. Player detection and identification followed by temporal segmentation of each scored point.

Following the presentation of an end-to-end architecture for autonomous analysis of broadcast badminton films, a pipeline based on off-the-shelf object identification, action recognition, and segmentation modules is built. Analytics for many sports rely on these modules, making their pipeline general for various sports, particularly racket sports (tennis, badminton, table tennis, etc.). Although these modules are trained, fine-tuned, and utilised individually, they have calculated different important and accessible metrics for higher-level analytics from each of these modules. The measurements may be computed or used differently for different sports, but the basic modules are seldom altered. This is since broadcast videos of many sports have similar issues. Rare short-term (e.g., deceit) and long-term (e.g., footwork around the court) techniques are inferred with varying degrees of confidence but are not automatically identified in our present method.

A robust fine-grained action identification approach would be required to detect a deception scheme that might mislead humans (players and annotators). On the other hand, predicting footwork necessitates long-term recall of game conditions. This study does not cover these parts of the analysis. Another difficult challenge is predicting a player's reaction or position. Because of the fast-paced nature of the games, sophisticated strategy, and various playing styles, it is extremely difficult for sports footage.

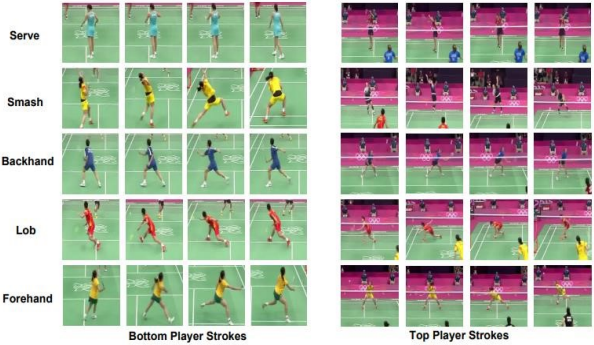

Figure 6. Representative "strokes" of bottom and top players for each class taken from our Badminton Olympic Dataset. The images have been automatically cropped using bounding boxes obtained from the player detection model. Top player appear smaller and have more complex background than the bottom player, therefore, are more difficult to detect and recognize strokes

### III. TECHNOLOGY

#### III.I. COMPONENTS OF SPORTS ANALYSIS

To create an AI coach, the necessary technology and tools for sports analysis are required. Following careful analysis, the functioning algorithms and tools that made the list are given below.

a) TrackNet
b) OpenPose
c) YOLOv3
d) IMU sensors
e) Opal sensor from APDM
f) Video recording system
g) 9-degree freedom wireless sensor
h) XiaoYu 2.0 Badminton sensor

After understanding the 'why' and 'how' and researching the many strategies available in this sector, describing the same in layperson's words is straightforward. TrackNet, a CNN, plots the shuttlecock trajectory on a heat map. The athlete's skeleton is then identified using OpenPose, which Carnegie University developed. YOLOv3 is used to detect the player's bounding box. The Opal sensor can track the players' motions systematically and precisely. Using the XiaoYu Badminton sensor, an intelligent racket will be an odd aid. It measures the speed of the racket as well as other factors such as handle pressure.

A video recording device captures the player from three 120-degree perspectives. It is done for the same shot to ensure uniformity. The specifics are provided below.

#### III.II.I. OPENPOSE

This represents the first real-time multi-person system to detect human body, hand jointly, and facial and foot key points on single images. There are, in total, 135 key points. Authored by the Carnegie University [6] [7] [8], it is used to map out the skeleton system.

It [6] is used for Body and Foot Estimation and constructing a 3-D Module for the body, face, and hands.

It [7] has features such as Functionality, 3-D real-time single person keypoint detection, Calibration toolbox, single person tracking, Input and Output for the training datasets. It is used for Body and Foot Estimation and constructing a 3-D Module for the body, face, and hands.

After processing the video using OpenPose, we map the skeletal system of the player. Advanced OpenPose feature allows localized mapping of different body parts. A whole body 3-D reconstruction and estimation could be done which would help with removing the noise in the data which has to be processed.

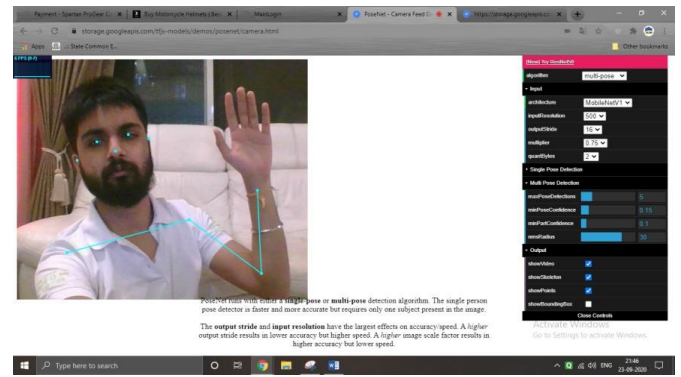

Figure 8. OpenPose depiction [9]

#### III.II.II TRACKNET

It is composed of a CNN followed by a de-convolutional (DeconvNet). It takes consecutive frames to generate a heat map indicating its position. The network parameter is the number of input frames. Input frame is considered the conventional CNN network. TrackNet with more than one input frame can improve the moving object tracking by learning the trajectory pattern. It is trained to generate the probability-like detection heat map having the exact resolution as input frames. Ground truth of heat map is amplified 2D Gaussian distribution located at the centre of the badminton. The ball coordinates are available in the labelled dataset, and the variance of the Gaussian distribution refers to the diameter of badminton images. The input of the proposed network can be some number of consecutive frames. The first 13 layers refer to the first 13 layers of VGG-16 for object classification. The 14-24 layers refer to DeconvNet for semantic segmentation. Upsampling is used to recover the information loss from the maximum pooling layers to achieve pixel-wise prediction. Maximum pooling layers and symmetric numbers of upsampling layers are used.

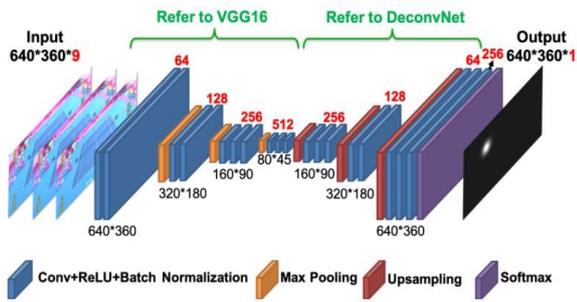

Figure 9. Architecture of TrackNet

### III.II.III. OPAL SENSOR FROM APDM

It is a research-grade wearable sensor designed for total control. It supports up to 24 Opals on a single network, can stream or log data, and has a separate Gyroscope and accelerometer. Stream to MATLAB, Java, and Python can access raw kinematic data and sync with other systems extremely effectively with a high sampling rate. The wireless synchronisation allows the athlete's natural movement to be recorded and real-time results to be obtained.

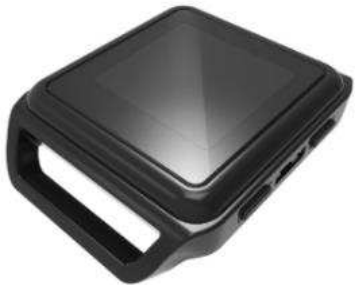

Figure 10. Opal Sensor [10]

### III.II.IV. YOLOv3

You Only Look Once is a cutting-edge, real-time object detection system in the long run. On COCO test-dev, a Pascal Titan Xt processes pictures at 30 frames per second and has a mAP of 57.9 %. YOLOv3 is highly quick and precise. In terms of mAP, IOU YOLOv3 is comparable to Focal Loss but around four times quicker. Furthermore, we may change the pace and precision from one to the other.

They used a single neural network to assess the whole image. This network separates the input into regions and predicts each bounding box and probabilities. The projected probability weighs these bounding boxes. Compared to classifier-based systems, this paradigm offers significant benefits. At test time, it examines the entire image; thus, the global context in the image impacts its predictions. It also predicts with a single network assessment instead of systems such as R-CNN, which take thousands for a single picture. This allows it to be 1000x quicker than R-CNN and 100x faster than Fast R-CNN.

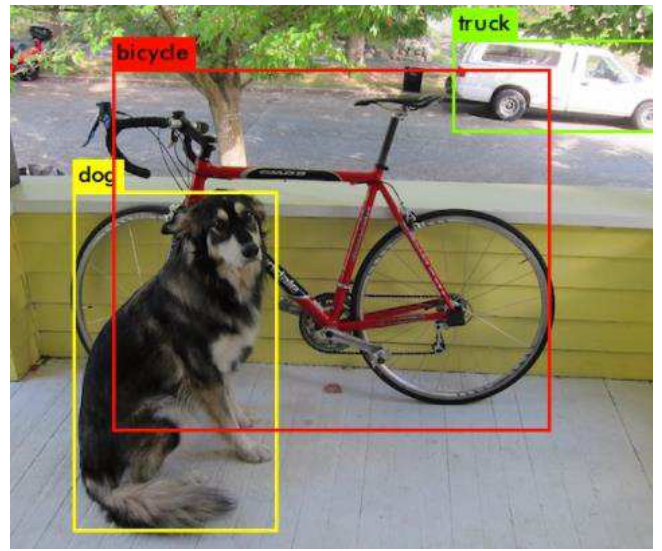

Figure 11. Prediction using YOLOv3 (localization) [11]

### IV. PROPOSED WORKING OF COMPONENTS

Let us consider each basic step of the concept, i.e., the flow of the procedure. It is very critical to understand the intricacies and not skip through.

*1)* Choose a topic player who is an amateur in badminton. Let us call this player – subject A.

*2)* Finalise a professional badminton player whose readily and adaptably footage is available on the world wide web or locally. Let us call this player – subject B.

*3)* The videos readily available of subject B can be downloaded from the internet. If subject B is available locally, then high-resolution videos of the subject can be taken as per convenience. The purpose of creating this dataset is to train the neural network.

*4)* The dataset has to be clean and not contain any noise. If we consider one particular class of shot, let us say – a 'Lift', then the video must pass these criteria. The player must be visible. Three videos must be taken of the same shot from three angles – 120 degrees from each adjacent camera. The player's hand, body, and leg movements must be visible.

*5)* After we get the dataset for training and testing the dataset, we call this dataset – Da (Dataset of subject A) and Db(Dataset of subject B).

*6)* After passing the video criteria, we use OpenPose and remove the unnecessary parts of both the Datasets – 'Da' and 'Db'. Thus we have remained with only the movements of the subject. Using a bounding box, we get the area in which the subject moves. Let us call this new dataset – 'Oda' (OpenPose Data of subject A) and 'Odb' (OpenPose Data of subject B).

*7)* For further examination of the shuttlecock and the leg movements of both the subjects, using TrackNet, it is possible to create a heatmap of the court during the match. Through this, we get the most probable regions that the shuttlecocks fall for a given class of shot from any particular area on the court. Using enough data, pattern analysis on the entire TrackNet dataset is possible.

*8)* We have the dataset and define the required parameters for training the model. The parameters include the tools and technologies defined.
   a. Setting different rotational limitations for a particular class of shot. This is done for both datasets. By setting the parameters of ODb for Oda, we see the range of inaccuracies and quantify these inaccuracies to calculate the accuracy of play for subject A.
   b. TrackNet data which is optional for pattern analysis of both the shuttlecock and the movements of the player.
   c. APDM Opal Sensor to measure the players' speed. Gait analysis is used to assess and treat individuals in sports biomechanics. Using these wearable sensors, it is possible to measure spatial/temporal gait and balance and full-body kinematics in unconstrained environments. Joint angle analysis while performing any movement on the court is instrumental in judging right from wrong.
   d. YOLOv3: OpenPose already offers the features of localization and bounding boxes but YOLOv3 does it at a much faster and efficient rate. It is possible to get a much higher accuracy if localization of the dataset is done before the creation of 'ODa' and 'ODb'.
   e. Inertial Measurement Unit. IMUs can be used to measure a variety of factors, including speed, direction, acceleration, specific force, angular rate. It is possible to modify the model on the basis of each players personalised usage.
   f. XiaoYu 2.0 Badminton sensor is the most advanced badminton sensor explicitly designed for the sport of badminton. Sports data is recorded instantaneously or delayed enabling a broader scope of improvement of subject A with advanced Gyroscope and Accelerometer technology to capture your racket's motions. Data is transferred to the device using Bluetooth. Speed, force, radian, and calories burned are all recorded. Six badminton motions are gathered for style analytics: "Smash," "Lift," "Clear," "Block," "Slice," and "Drive." Stroke training is also part of the package.

*9)* Setting the parameters, we train the model using a combination of the techniques specified in Literature Survey (II).

*10)* The testing dataset of ODb is fed into the model, and the result is the inefficiency assessment. The model would structure and produce this report after comparing the subject B to subject A and delivering the faults, inefficient, and incorrect plays executed.

*11)* Subject A should work on the identified mistakes. Slowly, the efficiency would rise, as would the model's accuracy, and the subject would become increasingly impossible to differentiate from subject B.

This is depicted in the flowchart below.

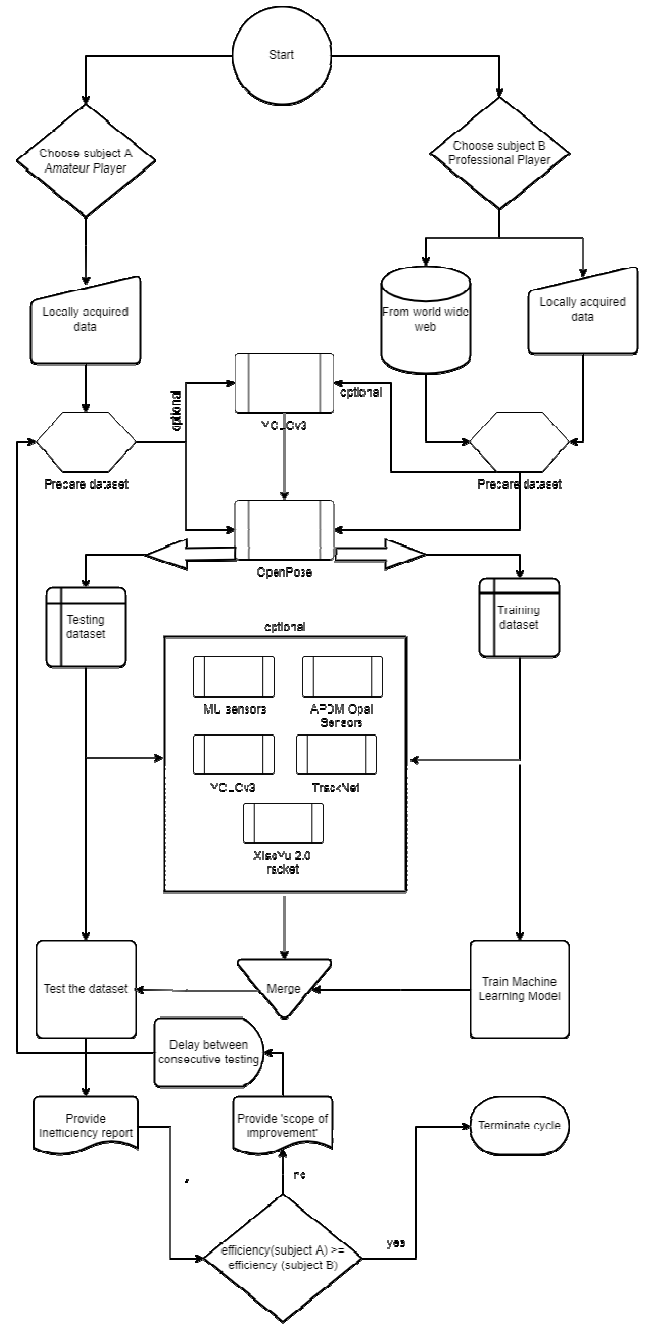

Figure 11. Flow chart of procedure

## V. Conclusion

By the end of this research paper, the definition of sports analysis is clear. The differences between the efficiency of a human coach and an AI coach coupled with the necessary improvements done at a constant rate shows that an AI coach is the future of analysing different sports. Also, after reviewing the numerous study publications, it is possible to infer that an AI coach would be a suitable alternative for a human coach. The reasons for this are compelling enough to warrant implementation, and by recognising that there can be no extension other than AI, it is also possible to deduce that human touch may be lacking in this area.

The players may use the report generated by the model to train themselves efficiently. The model's increased efficiency will assist the players in distinguishing themselves from the model players. But, as a result, having a human coach as your primary adviser is critical as studying and anticipating unprecedented events can only be avoided and used to the benefit of the athlete under the supervision of a professional human coach. The AI coach should be a route for the athlete to follow to improve.

## VI. Future work

Many alternative adaptations, tests, or experiments that might be conducted on the player to further customise and personalise it have been omitted. Future development might involve a more tailored approach to developing ideas for improving the game.

The things that influence a player's game are listed. Nutrition, sleep patterns, the best time to practise, injuries, a history of injuries and problems, weight class, height class, body type, and a variety of other physical and mental aspects that may influence a player's performance on the court are all elements to consider. The following ideas might be put to the test:

*a)* It will be interesting to see how the model behaves for various participants, grouped later based on their nationality. This might detect a trend and identify a player's optimal set of requirements to be near-perfect.

*b)* The model's accuracy is determined by the neural network's capacity to test the dataset provided by the user successfully. Any faulty dataset will render the model useless. To maximise the model's efficiency, the data presented to it must be clean. A kit including all necessary information might be offered to explain to the layman throughout the application's usage to eliminate data discrepancies. The package might include the required cameras, a handbook, and access to the application.

The primary future objective should be the real-time application of the concepts and technologies mentioned in the paper.